\documentclass[a4paper,11pt]{article}
\pdfoutput=1
\usepackage{jcappub}
\usepackage{amsmath, empheq}
\usepackage{amsfonts}
\usepackage{amssymb}
\usepackage{marvosym}
\usepackage{wasysym}
\usepackage[font=small,labelfont=bf]{caption}
\usepackage{subcaption}
\usepackage{fullpage}
\usepackage[]{graphicx}
\usepackage{hyperref}
\usepackage[english]{babel}
\usepackage{xcolor}
\usepackage{makeidx}

\numberwithin{equation}{section}

\newcommand{\be}{\begin{eqnarray}}
\newcommand{\ee}{\end{eqnarray}}

\hypersetup{citebordercolor=0 1 0, linkbordercolor=1 1 1,}
\title{\boldmath Dark matter in  $R+R^2$ cosmology}

\author[a,b]{E.V. Arbuzova,}
\author[b,c]{A. D. Dolgov,}
\author[b]{R. S. Singh}

\affiliation[a]{Department of Higher Mathematics, Dubna State University, \\Universitetskaya ul. 19, Dubna 141983, Russia}
\affiliation[b]{Department of Physics, Novosibirsk State University, \\Pirogova 2, Novosibirsk 630090, Russia}
\affiliation[c]{ITEP, Bol. Cheremushkinskaya 25, Moscow 117218, Russia}

\emailAdd{arbuzova@uni-dubna.ru}
\emailAdd{dolgov@fe.infn.it}
\emailAdd{akshalvat01@gmail.com}

\abstract{
Production of massive stable relics in $R^2$-modified gravity is considered. It is shown that 
the cosmological evolution and  kinetics of massive species differs significantly from those in 
the conventional cosmology. The results are applied to the lightest supersymmetric particles
and it is argued that they are viable candidates for the constituents of dark matter, if their mass
is about 1000 TeV. 
}

\begin{document}
\maketitle
\flushbottom

\pagenumbering{arabic}

\section{Introduction \label{s-Intro}}
Since the first indications to existence of Dark Matter (DM) by Kapteyn \cite{Kapteyn:1922zz}, Oort \cite{Oort32} and 
 %the discovery of Dark Matter (DM) 
  Zwicky~\cite{Zwicky:1933gu} in 1933 and later confirmation in 1970s \cite{Rubin:1970zza,Einasto-Nature-250,Ostriker:1974lna} 
  many theoretical models have been proposed to describe this elusive form of matter. 
  
An accepted property of  the DM particles is that they are electrically neutral\footnote{It is nevertheless possible that dark matter particles
have a tiny electric charge or even a "normal"  charge but very high mass, so the Thomson scattering is strongly suppressed. }
 since they don't scatter light, hence the 
name \textit{Dark} matter. Otherwise their properties are practically unknown.
This opened possibilities for innumerable particles to be DM candidates.  
  
A natural, and formerly very popular, candidate for dark matter particle is the lightest supersymmetric  particle (LSP) which should be stable if the so-called R-parity is conserved. The latest reviews on SUSY dark matter, and not only, can be found in 
Refs.~\cite{Catena:2013pka,Gelmini:2015zpa,Lisanti:2016jxe,Slatyer:2017sev,Cline:2018fuq}.

%However, thanks to the precise cosmic microwave background based calculation of relic DM density and the sensitive underground experiments for DM scattering[Arbey], which provides powerful constraints for many DM candidates.  This brings us to the topic of supersymmetry(SUSY), which provides several DM candidate, collectively called Lightest Supersymmetric Particles(LSPs). The popular DM candidate LSP are the lightest neutralino, the gravitino or the lightest sneutrino, given the model.\\~\\
%Later, many studies showed that sneutrino in the Minimal-Supersymmetric model(MSSM) cannot be a DM candidate because of its strong coupling with the Z-bosons, resulting in a  large annihilation cross-section, thus making it easy to detect but no observations have been made. Consequently, many extended models were proposed to reinstate sneutrino as a DM candidate.  A detailed reading on all SUSY dark matter can be found in Catena et. al.~\cite{catena}.

An extensive search for the low energy supersymmetry performed at LHC led to negative results. Thus, if supersymmetry exists, its characteristic energy scale should be, roughly speaking, higher than 10 TeV. The cosmological energy density of LSPs is proportional to their mass squared, 
$\rho_{LSP} \sim m_{LSP}^2$, and for $m_{LSP} \sim 1$ Tev $\rho_{LSP}$ is of the order of the observed energy density of the universe.
Correspondingly for larger masses such particles would overclose  the universe. This  unfortunate 
circumstance excludes LSPs as dark matter particles in the conventional cosmology. 

There are several attempts in the literature to save  supersymmetric dark matter by modifying the cosmological
scenarios of LSP production in such a way that the relic density of heavy LSP would be significantly suppressed.
For example in the paper~\cite{kane-non-therm} a detailed study of non-thermal production of heavy relics is performed.
Recently in Ref.~\cite{dreese} a specific scenario has been studied, which is based on the assumption
that after the freezing of LSP the universe was matter dominated and this epoch
transformed into the radiation dominated stage with low reheating temperature.  Similar idea was discussed earlier in 
paper~\cite{DNN}, where it was assumed that at some early stage the universe might be dominated by primordial black holes
which created the necessary amount of entropy to dilute the heavy particle relics.

The $R^2$ inflation was generalized to supergravity in the series of 
papers~\cite{Ketov:2010qz,sg2,Ketov:2018uel} and references therein.
In particular in Ref.~\cite{sg2} a scenario with superheavy gravitino, which 
may be a viable candidate for  dark matter particle, was considered. The mechanism
considered there is different from that proposed below.

In this paper we show that in $(R+R^2)$-gravity the energy density of LSPs may be much lower and so it reopens for them the chance  to be the dark matter. This possibility was mentioned in our previous paper~\cite{Arbuzova:2018ydn} but we didn't present there any calculations, which are done in the present work.

The paper is organized as follows. 
In Sec. \ref{s-R2-GR} we summarise the essential features of cosmological evolution of the energy density of ordinary matter in Starobinsky inflationary scenario \cite{Gurovich:1979xg,Starobinsky:1980te,Starobinsky_1982} and compare them with that in General Relativity (GR).   
In Section \ref{s-kinetic-scalars} the cosmological density of massive stable relics, in particular of LSP, is calculated for the case of dominant 
production of scalar particles by the oscillating curvature $R(t)$ (in other words, by the scalaron decay). 
It is assumed in this Section that the scalar particles are minimally coupled to curvature.
In the following Sec.~\ref{s-decay-fermion} the same is done for the predominant scalaron decays into spin 1/2 fermions
or into scalars with conformal coupling to curvature. 
The results of sections \ref{s-kinetic-scalars} and \ref{s-decay-fermion}  are grossly different. In Sec.~\ref{s-gauge-anomaly} 
comments on gauge boson production due to conformal anomaly are presented.
Lastly we conclude.

%we obtain the kinetics for the particle $X$ in both GR and $R^2$ regime and it was found that indeed the kinetics in $R^2$ theory are different from that of GR. In section\ref{decay-fermion} we discuss the decay of scalaron into fermions and obtain values for the reheating temperature($T_h$), the thermal equilibrium condition and the freezing temperature($T_f$). In Sec.[\ref{numberdensity}], we obtain the ratio of number density of X-particle to that of photons and de

\section{Evolution of matter density: $R^2$-gravity versus General Relativity \label{s-R2-GR}} 

The cosmological evolution in $(R+R^2)$ theory was studied in the paper~\cite{Arbuzova:2011fu} and recently 
in our work~\cite{Arbuzova:2018ydn}.
 According to Ref.~\cite{Arbuzova:2018ydn}, the cosmological evolution in $R^2$-theory can be separated into four distinct epochs.  Firstly, there was the inflationary stage, when the curvature scalar was sufficiently large, and the universe expanded exponentially with slowly decreasing $R(t)$. The second epoch began, when $R(t)$ approached zero and started to oscillate around it as
\be
%&&h= \frac{2}{3\tau} \left[1+ \sin(\tau + \theta )\right] \label{hsol}, \\
R=-\frac{4m_R\cos(m_Rt+ \theta)}{t}, \label{rsol}
\ee
where $m_R$ is the scalaron mass taken as $3 \times 10^{13}$ GeV.
At this stage the Hubble parameter has the form
\be
H= \frac{2}{3t} \left[1+ \sin(m_Rt + \theta )\right]. \label{hsol}
\ee
The oscillations of $R$ gave rise to the particle production, but the energy density of the produced particles was negligible and had no noticeable impact on the cosmological evolution. So this period can be called scalaron dominated regime. It was followed by the transition period from scalaron domination to domination of the usual (relativistic) matter. 
Lastly, after complete decay of the scalaron we arrive to the conventional cosmology governed by General Relativity. 

Here we are mostly interested in the scalaron dominated period, during which cosmology and particle kinetics essentially differ from those in GR.
The energy density of the produced  particles depends upon the form of their coupling to  curvature $R(t)$. Because of the Parker theorem~\cite{Parker:1968mv}, production of massless particles in FLRW-metric is absent in conformally invariant theory. 
Massless scalar field with minimal coupling to gravity, as known, is not conformally invariant. 
The width of the decay into two scalars does not vanish in the limit of zero mass, see e.g.~\cite{Arbuzova:2018ydn}, and is equal to:
\be
\Gamma_s = \frac{m_R^3}{48m_{Pl}^2},
\label{Gamma-s}
\ee
where $m_{Pl}$ is the Planck mass.
Correspondingly, the energy density of massless scalars, $\rho_{s}$, created by the oscillating $R(t)$, is: 
\be
\rho_{s} = \frac{m_R^3}{120 \pi t}. 
\label{rho-s}
\ee
Massless fermions are conformally invariant and so only massive fermions can be produced with probability proportional to the square of their mass, $m_f$. The width of the scalaron decay into pair of fermions
is~\cite{Gorbunov:2010bn}: 
\be
\Gamma_f = \frac{ m_R m_f^2 }{48 m_{Pl}^2 },
\label{Gamma-f}
\ee
%where $m_{Pl}$ is the Planck mass and $g_f$ is the number of different types of fermions. 
Decay probability is dominated by the heaviest fermion. 
More accurately we should take sum over all fermionic species. 
%but we assume that all fermions have equal mass, $m_f$. 
Hence the energy density of the produced relativistic fermions is equal to 
\be 
\rho_f = \frac{ m_R m_f^2}{120 \pi t}.
\label{rho-f}
\ee
 Similar suppression factor proportional to the boson mass squared appears for scalar particles with conformal coupling to gravity.

 Normally the energy density of relativistic matter drops down as $1/a^4(t)$, where $a(t)$ is the the cosmological scale factor, which at the described stage rises as $a(t) \sim t^{(2/3)}$. So naively one would expect that 
$\rho \sim 1/t^{(8/3)}$. 
 However, the laws (\ref{rho-s}) and  (\ref{rho-f}) 
 demonstrate much slower decrease of the energy density of matter, which is ensured by  the flux of energy from the scalaron decay.  

Expressions (\ref{rho-s}) and  (\ref{rho-f}) can be compared with the energy density of matter in the 
standard GR cosmology: %which drops down as ${1}/{t^2}$:
\be
\rho_{GR} = \frac{3H^2m_{Pl}^2}{8 \pi} = \frac{3 m_{Pl}^2}{ 32 \pi t^2}.
 \label{rho-GR}
\ee
%where $m_{Pl}$ is the Planck mass.

%We assume that matter reached thermal equilibrium state and 
Let us derive
equations connecting temperature with time for different expressions for energy density of matter  \eqref{rho-s}, \eqref{rho-f}, 
and \eqref{rho-GR}. 
Assuming that the equilibrium with 
temperature $T$ is established, we estimate the particle reaction rate as
\be
\Gamma_{scat} \sim \alpha^2 \beta_{scat} T,
\label{Gamma-par}
\ee
where $\alpha$ is the coupling constant of the particle interactions, typically $\alpha \sim 10^{-2}$, and $\beta_{scat}$ is the number of scattering  channels, $ \beta_{scat} \sim 100$. 
Equilibrium is enforced if $\Gamma_{scat} > H \sim 1/t$ or $\alpha^2 \beta_{scat} T t > 1$. The energy density of relativistic
matter in thermal equilibrium is expressed through the temperature as: 
\be
\rho_{therm} = \frac{\pi^2 g_*}{30}\, T^4 
\label{rho-therm}
\ee 
where $g_*$ is the number of relativistic species in the plasma. We take $g_* \sim 100$.

Using equations \eqref{rho-s} and \eqref{rho-therm}, we find the equilibrium condition for the case of scalaron decay into a pair of massless scalars:
\be 
\left( \alpha^2 \beta_{scat}  Tt\right)_{s} =\frac{ \alpha^2 \beta_{scat}}{4\pi^3 g_*} \left(\frac{m_R}{T}\right)^3 \approx  8\cdot 10^{-7}  \left(\frac{m_R}{T}\right)^3 > 1. 
\label{Teq-s}
\ee  
Analogously from \eqref{rho-f} and \eqref{rho-therm} follows that thermal equilibrium in the case of the decay into a pair of fermions is established, when
 \be 
\left( \alpha^2 \beta_{scat} Tt\right)_{f} =\frac{ \alpha^2 \beta_{scat}}{4\pi^3 g_*} \frac{ m_R m_f^2}{T^3} \approx
  8\cdot 10^{-7} \, \frac{ m_R m_f^2}{T^3} > 1. 
\label{Teq-f}
\ee  
On the other hand, for the GR-cosmology it follows from Eqs.~\eqref{rho-GR} and \eqref{rho-therm} that the equilibrium is established when:
\be 
\left(\alpha^2  \beta_{scat} Tt\right)_{GR} =\alpha^2 \beta_{scat} \left(\frac{90}{32\pi^3 g_*}\right)^{1/2} \frac{m_{Pl}}{T}
\approx 3\cdot 10^{-4} \,\frac{m_{Pl}}{T} > 1. 
\label{Tt-GR}
\ee
%Here $\alpha$ is  the coupling constant which is usually $\alpha \sim 10^{-2}$ and $g_*$ is the number of relativistic species in the plasma, taken to be $g_* \sim 100$.

Let us estimate now the so-called heating temperature, that is the temperature of the cosmological plasma after complete decay of the scalaron. 
It can be estimated
from the expression for the energy density of matter at the time moment equal to the inverse decay width $t_d=1/\Gamma$. For the decay into scalars $\Gamma $ is given by Eq.~\eqref{Gamma-s} and 
\be
\rho_s(t_d=1/\Gamma_s)= \frac{m_R^3}{120 \pi t_d}=\frac{m_R^6}{5760 \pi m_{Pl}^2}=\frac{\pi^2}{30}\,g_*T_{hs}^4\,.
\label{rho-s-d}
\ee
Hence the heating temperature for the dominant decay of the scalaron into scalar particles is equal to:
\be
T_{hs} \approx \frac{m_R}{(192\pi^3)^{1/4}}\,\left({\frac{m_R}{m_{Pl} }}\right)^{1/2}.
\label{T-hs}
\ee
For $m_R = 3 \times 10^{13}$ Gev $T_{hs} \approx 6\times 10^8$ GeV. %, which is close to other estimates presented in the literature.

Analogously, using Eqs. \eqref{rho-f} and \eqref{Gamma-f},  we find the temperature of the universe heating for the case of scalaron 
decay into fermions:
\be
T_{hf}  = \frac{ 1 } 
{(192 \pi^3 g_*)^{1/4} }
\left( \frac{m_R}{m_{Pl}} \right)^{1/2} \, m_f \approx 5.7\cdot 10^{-5} m_f.
\label{T-hf}
\ee

In both cases the heating temperature is considerably lower than  the  temperature at which thermal equilibrium is established, see 
Eqs. \eqref{Teq-s} and \eqref{Teq-f}.

\section{LSP density for the scalaron decay into scalars \label{s-kinetic-scalars} }

The freezing of massive particle species, $X$, is governed by the following equation:
\be
\dot n_X + 3H n_X = -\langle \sigma_{ann} v \rangle \left( n_X^2 - n^2_{eq} \right),
\label{dot-n-X}
\ee
where $n_X$ is a number density of particles $X$, $\sigma_{ann}$ is their annihilation cross-section, $v$ is the center-of-mass velocity.  
The angle brackets mean thermal averaging over the medium. For annihilation of the non-relativistic particles the averaging is not essential, i.e.
 $\langle \sigma_{ann} v \rangle= \sigma_{ann} v$.
The equilibrium number density of $X$-particles, $n_{eq}$,  is  given by the expression 
\be
n_{eq} = g_s \left(\frac{M_X T}{2\pi}\right)^{3/2} e^{-M_X/T},
\label{n-eq}
\ee
where $g_s$ is the number of spin states of $X$-particles and $M_X$ is their mass.

The annihilation cross-section can be estimated as
\be
\sigma_{ann} v = \frac{\alpha^2 \beta_{ann}}{M_X^2},
\label{sigma-ann}
\ee 
where $\beta_{ann}$ is a numerical parameter proportional to the number of open annihilation channels, it can be of order of ten or even larger;
 $\alpha$ is a coupling constant. In supersymmetric theories $\alpha \sim 0.01$.

This equation was derived in 1965 by Zeldovich \cite{Zeldovic:1965} and collaborates \cite{Zeldovic:1965rys,Zeldovic:1965UFN}. Twelve years later equation \eqref{dot-n-X} was applied to determination of cosmological density of heavy neutral leptons \cite{Lee:1977ua,Vysotsky:1977pe}.
Presently this equation is called the Lee-Weinberg equation, though justly  it should be  called the Zeldovich equation.  

In our case an additional term describing $X$-particle production by $R(t)$ should be included into Eq.~\eqref{dot-n-X}. However we assume that this channel is suppressed in comparison with inverse annihilation of light particles into $X\bar X$-pair. 

We do not specify which precisely supersymmetric particle is the lightest (it  can be, e.g., sneutrino, neutralino or gravitino). We only need the value of its mass, $M_X$, and the magnitude of the annihilation cross-section \eqref{sigma-ann}.   

We assume that the plasma is thermalised that is the temperature satisfies the condition \eqref{Teq-s}.

In what follows we perform calculations in parallel for $R^2$-cosmology and for the classical  GR-cosmology. 

During the radiation dominated stage in GR frameworks we have:
 \be 
H= 1/(2t),\,\,\,\, a(t) \sim t^{1/2}.
\label{H-GR}
\ee
Equating the critical energy density \eqref{rho-GR} to the energy density of relativistic plasma with temperature $T$ \eqref{rho-therm}
\be 
\rho_{GR} = \frac{3 m_{Pl}^2}{32\pi t^2} = \frac{\pi^2 g_* T^4}{30},
\label{rho-GR-T}
\ee
we find
\be
t T^2 = \left(\frac{ 90}{32 \pi^3 g_*}\right)^{1/2} \,m_{Pl} = const.
\label{t-T2}
\ee
Correspondingly
\be 
\frac{\dot T}{T} = - \frac{1}{2t}.
\label{dot-T-GR}
\ee
Analogously, using Eqs.~\eqref{rho-s} and \eqref{rho-therm}
\be 
\rho_{s} = \frac{m_R^3}{120\pi t} = \frac{\pi^2 g_* T^4}{30},
\label{rho-s-T}
\ee
we obtain:
\be
t T^4 = \frac{ m_R^3}{4 \pi^3 g_*}= const
\label{t-T4}
\ee
and
\be 
\frac{\dot T}{T} = - \frac{1}{4t}.
\label{dot-T-R2}
\ee

To eliminate $3Hn$ term in Eq.~\eqref{dot-n-X} let us introduce the new function, $f$, according to the relation
\be 
n_X = n_{in} \left(\frac{a_{in}}{a}\right)^3 f  ,
\label{n-normal}
\ee
where  $n_{in}$ is the value of $X$-particle density at $a=a_{in}$ and $T_{in}=M_X$, so the $X$-particles can be considered as relativistic and
thus 
\be
n_{in} = 0.12 g_s T^3_{in} = 0.12 g_s M_X^3. 
\label{n-in}
\ee
It is also convenient to introduce new variable $x = M_X/T$.

As we see in what follows, the final result does not depend upon $n_{in}$ and $T_{in}$.

In the case of the conventional GR-cosmology using Eqs. \eqref{t-T2} and \eqref{dot-T-GR}, we arrive to the equation:
\be
\frac{df}{dx} = - m_{Pl} M_X \sigma v \left( \frac{45}{4 \pi^3 g_*}\right)^{1/2} \, \left(\frac{a_{in}}{a}\right)^3 \frac{n_{in}}{ T^3}\,
\frac{\left( f^2 - f_{eq}^2 \right)}{x^2}.
\label{df-dx-GR}
\ee
Analogously,  in $R^2$ theory using Eqs. \eqref{t-T4} and \eqref{dot-T-R2}, we obtain
\be
\frac{df}{dx} = - \sigma v \, \frac{m_R^3}{\pi^3 g_* M_X}\, \left(\frac{a_{in}}{a}\right)^3 \frac{n_{in}}{ T^3}\,
\left( f^2 - f_{eq}^2 \right).
\label{df-dx-R2}
\ee

%Before solving Eqs.~(\ref{df-dx-GR}) and (\ref{df-dx-R2})
We need to determine the values of the products  $({a_{in}^3 n_{in}})/({a^3 T^3})$,
which are very much different in GR and $R^2$ theories. During the GR regime the product $a T $ is almost constant, up to the corrections
induced by the heating of  the plasma by the massive particles annihilation when the temperature drops below their masses. That's
why the number density of $X$-particles calculated in GR are
normalized not to the photon density but to the entropy density which is conserved in
the comoving volume. Up to this factor the coefficient $ ({a_{in}^3 n_{in}})/({a^3 T^3})$ in the GR regime can be taken as unity.

%It is worth noting at this point that in GR $a T \approx const$ up to entropy release factor due to massive particle annihilation,
%while 
In $R^2$ theory $T \sim t^{-1/4}$, $a\sim t^{2/3}$, and $a^3 T^3 \sim 1/T^5$,  %Such a difference is induced by a constant
%heating of the primeval plasma by the scalaron decay in $R^2$-theory. 
%In $R^2$-regime before the scalaron decay this ratio essentially changes. 
so the ratio $(a_{in}/a(t))^3$ can be estimated as
\be
\left(\frac{a_{in}}{a(t)}\right)^3 = \left(\frac{t_{in}}{t}\right)^2 = \left(\frac{T_{in}}{T}\right)^{ - 8} =\frac{1}{x^8}.
\label{a-in-over-a}
\ee
The initial number density of $X$-particles is taken according to Eq.~\eqref{n-in} and so 
%as the equilibrium number density if relativistic particles
%with $g_s$-number of spin states and temperature $T_{in} = M_X$. So
\be
\frac{n_{in}}{T^3} = 0.12 g_s \left(\frac{M_X}{T}\right)^3 =0.12 g_s x^3.
\label{n-over-T3}
\ee
Since the density of $X$-particles is given by eq.~(\ref{n-normal}),
its ratio to the number density of photons, $n_\gamma = 0.24 T^3 $, is equal to:
\be
\frac{n_X}{n_\gamma} = \frac{g_s}{2}\,\frac{f}{x^5}.
\label{n-over-n-gamma}
\ee

The equation \eqref{df-dx-R2}  governing the evolution of $X$-particles in the scalaron dominated regime is transformed to
\be
\frac{df}{dx} =  - \frac{0.12 g_s \alpha^2 \beta_{ann}}{\pi^3 g_*} \left(\frac{m_R}{M_X}\right)^3 \, \frac{f^2 - f_{eq}^2}{x^5} 
\equiv - {Q_s}\, \frac{f^2 - f_{eq}^2}{x^5},
\label{df-dx-R2-2}
\ee
where 
\be
 Q_s =  \frac{0.12\, g_s\, \alpha ^2 \beta_{ann}}{\pi ^3 g_*} \left( \frac{m_R}{M_X}\right)^3.
\label{tilde-Q}
\ee
%does not depend on $x$.

Since the coefficient $Q_s$   in front  of $(f^2 - f^2_{eq})$ in Eq.~(\ref{df-dx-R2-2})
is normally huge, then  initially the solution is close to the equilibrium one:
\be 
f = f_{eq} (1 + \delta)
\label{r=req}
\ee
with $\delta$ equal to
\be
\delta =  - \frac{x^5}{2  Q_s f_{eq}^2} \frac{df_{eq}}{dx} \approx  \frac{x^5}{ 2  Q_s f_{eq}}.
\label{delta}
\ee
Note, that $df_{eq}/dx \approx - f_{eq}$ for large $x$ .This solution is valid till $\delta$ remains small, $\delta \leq 1$.
According to Eqs.~\eqref{n-eq}, \eqref{n-normal}, and \eqref{a-in-over-a}
\be
f_{eq} =\frac{1}{0.12} \left( \frac{x}{2\pi} \right)^{3/2} e^{-x} x^5.
\label{f-eq-2}
\ee
 and according to Eqs.~\eqref{delta} and \eqref{a-in-over-a} we have:
\be
\delta =  \frac{x^5}{ 2  Q_s f_{eq} }=  \frac{0.06 }{ Q_s} \left( \frac{x}{2\pi} \right)^{-3/2} e^{x}, 
\label{delta-1}
\ee
where $Q_s$ is given by Eq.~\eqref{tilde-Q}.
%\be
%Q^* = -\frac{2\,\tilde Q }{0.12 (2\pi)^{3/2}} = 
%\frac{ 2\,g_s\, \alpha ^2 \beta}{(2\pi)^{3/2} \, \pi ^3 g_*} \left( \frac{m}{M_X}\right)^3 
%\label{Q-star}
%\ee
The deviation from equilibrium becomes of order of unity, or $\delta =1$, 
 at the so-called freezing temperature $T_{fr}$ or at $x_{fr}$, which  
is approximately:
\be
x_{fr} \approx \ln Q_s +\frac{3}{2}\ln(\ln Q_s) - \frac{3}{2} \ln (2 \pi) + \ln 0.06 
\approx \ln Q_s +\frac{3}{2}\ln(\ln Q_s) - 5.7 \,.
\label{x-f}
\ee
Since  $Q_s \gg 1$, then $x_{fr}$  is also large, typically $x_{fr} \sim (10-100)$ depending upon the interaction strength.

After $x$ becomes larger than $x_{fr}$, $f^2_{eq}$ can be neglected in comparison to  $f^2$ and equation (\ref{df-dx-R2-2}) with the initial condition 
$f=f_{fr}$ at $x=x_{fr}$ is simply 
integrated giving the asymptotic result at $x \to \infty$:
\be
f_{fin} = \frac{f_{fr}}{1+ \frac{  Q_s f_{fr}}{4}\left( \frac{1}{x_{fr}^4} - \frac{1}{x^4}\right)}
 \rightarrow \frac{ 4 x_{fr}^4}{  Q_s}.
\label{f-fin}
\ee
The last asymptotic limit is valid when $x\gg x_{fr}$ and $Q_s f_{fr}/(4 x_{fr}^4) >1$. It  is fulfilled because $x_{fr} \sim \ln  Q_s$
is sufficiently large and 
%it can be so because  
according to Eq.~(\ref{f-eq-2})  
\be
%$f_f \sim x_f^{3/2+ 5}/ Q$.
\frac{Q_s f_{fr}}{4 x_{fr}^4} \sim x_{fr}^{5/2} \gg 1.
\label{Q-f-f}
\ee

Thus $f_{fr}$ tends to a constant value, $f_{fin}$ \eqref{f-fin}, but, according to Eq.~(\ref{n-over-n-gamma}), the ratio of the number densities of $X$-particles 
with respect to photons drops down strongly, as $1/x^5$, in contrast to the analogous ratio in GR regime. This decrease is 
induced by the rise of the density of relativistic species created by the scalaron decay. This drop continues till $\Gamma t \sim 1$,
when scalaron field disappears and the cosmology returns to the usual GR one. 
It happens at the temperature given by Eq.~\eqref{T-hs}. 
This temperature should be compared with the temperature of  the establishment of thermal equilibrium $T_{eq} \approx 2\times 10^{-3} m_R$, as follows from Eq.~\eqref{Teq-s}.

Our results are valid if $T_{eq} > M_X > T_{hs}$. However, if the condition $T_{eq} > M_X$  is not fulfilled, 
the cosmological number density of heavy massive particles with masses larger than $T_{hs}$ would still be suppressed 
and even stronger than in the case of $T_{eq} > M_X $.

So, to summarise, the ratio of the number densities of $X$-particles to that of photons drops  down at $R^2$ dominated regime as $1/x^5$, 
see Eq.~(\ref{n-over-n-gamma}), but after the complete decay of the scalaron this ratio remains essentially constant, as it is usual
in GR.  So the present day ratio $n_X/n_\gamma$ can be estimated as the value of this ratio at $T=T_{hs}$ (\ref{T-hs}), i.e. at 
$x$ equal to:
\be
x_{hs} = (192\pi^3)^{1/4}  \left(\frac{M_X}{m_R}\right)\,\left({\frac{m_{Pl}}{m_{R} }}\right)^{1/2}.
\label{x-decay}
\ee
Using Eq.~(\ref{f-fin}), \eqref{tilde-Q} and (\ref{n-over-n-gamma}), we estimate the number density of the $X$-particles at the present time as
\be
\left(\frac{n_X}{ n_\gamma }\right)_{now} = \frac{2 g_s x_{fr}^4}{Q_s x_d^5} = 
\frac{\pi^3 g_*  x_{fr}^4 }{ 0.06  (192 \pi^3)^{5/4} \alpha^2 \beta_{ann}}\, \left(\frac{m_R}{M_X}\right)^2
\left(\frac{m_{R}}{m_{Pl}}\right)^{5/2},
\label{n-X-n-gamma-today}
\ee
where $x_{fr}$ is determined by Eq.~\eqref{x-f}.

Taking $g_* =100$, $\alpha = 0,01$, $\beta_{ann} =10$, $m_R = 3\times 10^{13} $ GeV, and $n_\gamma = 412$ /cm$^3$ we find for
the present day energy density of the $X$-particles:
\be
\rho_X = M_X n_\gamma f_{fin} \approx  1.7\times 10^8 \left(\frac{10^{10} {\rm {Gev} }}{M_X} \right) \,\rm{keV/cm^3}.
%\left(\frac{M_X}{10^{10} {\rm GeV}}\right)^8\,{\rm GeV/cm^3}
\label{rho-X}
\ee
This is to be compared with the observed energy density of dark matter $\rho_{DM} \approx 1$ keV/cm$^3$.
We see that $X$-particles must have huge mass, much higher than $m_R$ to make reasonable dark matter density.
However, if $M_X > m_R$, the decay of the scalaron into $X\bar X$-channel would be strongly suppressed and such LSP with the
mass slightly larger than $m_R$ 
%with the mass $M_X \sim 2 \times 10^9 $ GeV 
could successfully make the cosmological dark matter. We will not further pursue this possibility here but turn in the next section to 
%the scalaron decays into 
LSP being a fermion or conformally coupled scalar.

\section{Decay into fermions {\bf or conformal scalars}}\label{s-decay-fermion}

Let us assume now that scalaron decays only to fermions or to conformally coupled scalars. 
If the bosons are coupled to curvature as 
%The bosons $\phi$ are conformally coupled to curvature, as 
$\xi R \phi^2 $ with $\xi = 1/6$, they are conformally invariant and are not produced if their mass is zero. The probability of production of both bosons and fermions is proportional to their mass squared.  
In what follows we confine ourselves to consideration of fermions only, because there is no essential difference between bosons and fermions. 
%that the lightest supersymmetric particle is a fermion. In this case 
The width of the scalaron decay into a pair of fermions is given by Eq.~\eqref{Gamma-f}. 
The largest contribution into the cosmological energy density at scalaron dominated regime is presented by the 
 decay into heaviest fermion species.

We assume that the mass of the lightest supersymmetric particle is considerably smaller than the masses of the other decay products, $m_X < m_f$, at least as $m_X \lesssim 0.1 m_f$. Then the direct production of $X$-particles by $R(t)$  can  be neglected.  In such a case
LSP are dominantly produced by the secondary reactions in the plasma, which was  created by the scalaron production of heavier particles.

Using expression \eqref{T-hf} for the temperature of the universe heating after the scalaron decay, we find  
\be 
x_{hf} \equiv \frac{m_X}{T_{hf}} \approx 1.75\cdot 10^4 \, \frac{ m_X}{m_f}.
\label{x-h-f}
\ee
%The numerical values are taken for $ m_R = 3\cdot 10^{13} $ GeV  and $ g_f = g_* = 100 $.

According to Eq.~\eqref{Teq-f} cosmic plasma thermalised at temperatures below 
\be 
T_{eqf} \approx 10^{-2} m_f \left(\frac{m_R}{m_f}\right)^{1/3}.
\ee

The time-temperature dependence, as follows from Eqs.~\eqref{rho-f} and \eqref{rho-therm}, is:
\be 
t T^4 = \frac{m_R m_f^2}{4\pi^3 g_*}.
\label{t-T4-f}
\ee
 
Kinetic equation for freezing of fermionic species can be solved in complete analogy with what was done in the previous section. 
The relative number density, $f$, is defined by the same relation \eqref{n-normal} and the kinetic equation for $f$ 
has the same form:

\be
\frac{df}{dx} = % - \frac{0.12 g_s \alpha^2 \beta_{ann}}{\pi^3 g_*} \left(\frac{m_R}{M_X}\right)^3 \, \frac{f^2 - f_{eq}^2}{x^5} 
 - {Q_f}\, \frac{f^2 - f_{eq}^2}{x^5} ,
\label{df-dx-R2-2-f}
\ee
where 
\be
 Q_f =  \frac{\alpha ^2 \beta_{ann}}{\pi ^3 g_*} \, \frac{m_R m_f^2}{m_X^6} \, n_{in},
\label{Q-f}
\ee
and $n_{in}=0.09 g_s m_X^3$ is the initial number density of $X$-particles at the temperatures $T \sim m_X$.

%Eq.~(\ref{df-dx-R2-2-f}) can be solved in the standard way. It is assumed that initially $r$ is close to its equilibrium value, This assumption is
%well justified. So we take
We take as in Sec. \ref{s-kinetic-scalars}
\be 
f \approx f_{eq} (1 +\delta),
\label{f-delta-f}
\ee
where %by assumption $\delta < 1$. It is enforced by a large factor $Q$. The solution in this case is pretty evident:
\be
\delta \approx \frac{x^5}{2Q_f f_{eq}}.
\label{delta-f}
\ee
The equilibrium relative number density, $f_{eq}$, is slightly different from the similar quantity for bosons \eqref{f-eq-2} and is equal to:
\be
f_{eq} =\frac{1}{0.09} \left( \frac{x}{2\pi} \right)^{3/2} e^{-x} x^5.
\label{f-eq-2-f}
\ee

The  freezing temperature is defined by
\be
x_{fr} \approx \ln Q_f +\frac{3}{2}\ln(\ln Q_f) - \frac{3}{2} \ln (2 \pi) + \ln 0.045 
\approx \ln Q_f +\frac{3}{2}\ln(\ln Q_f) - 5.86 \,,
\label{x-f-f}
\ee
%\textbf{check the equation.}
and the so called frozen value of $f$ is equal to
\be
f_{fr} = \frac{x_{fr}^5} {Q_f}.
\label{f-f}
\ee
Using expression (\ref{Q-f}) for $Q_f$ we find that the frozen number density of $X$-particles, i.e. taken at $T=T_{fr}=m_X/x_{fr}$ is
\be 
n_{Xfr} = \frac{\pi^3 g_*}{\alpha^2 \beta_{ann} \ln^3 Q_f}\,\frac{m_X^6}{m_R m_f^2}.
\label{n-f}
\ee

Some additional burning of $X$-particles takes place during the period, when 
 ${ f > f_{eq} }$ and  equation (\ref{df-dx-R2-2-f}) is
simplified to:
\be
\frac{df}{dx} = -Q_f\, \frac{f^2 }{x^5} .
\label{dr-dx-asympt}
\ee
The solution of this equation with the initial condition $f(x_{fr}) = f_{fr}$
for the asymptotic value at large $x\gg x_{fr}$ is trivially found:
\be
f_{fin} = \frac{f_{fr}}{1+ \frac{  Q_f f_{fr}}{4}\left( \frac{1}{x_{fr}^4} - \frac{1}{x^4}\right)}
 \rightarrow \frac{ 4 x_{fr}^4}{  Q_f}.
\label{f-fin-f}
\ee

This would be the asymptotic value of the relative number density of the heavy stable relics in the standard approach. 
However, $n_X$ does not drop down as $1/a^3$, but much faster due to the extra heating of plasma by the scalaron decay,
which does not create $X$-particles if their coupling to the scalaron is sufficiently weak, as it is assumed above.
One should, however, remember that there exists a continuous production  of heavier fermions, which as it is mentioned
in the second paragraph of this Section, is much stronger than the direct production of LSP, i.e. of the $X$-particles. However, 
the heavy fermions $f$ are produced with huge energy $E_f \sim m_R/2$. This energy is thermalized and is transformed into the
energy of relativistic species producing $m_R/T$ relativistic particles per one $X$ particles created by the $f$ decays.
Moreover, some heavy fermions could annihilate without creation of $X$ particles, but this effect is rather weak, even with
an account of relativistic delay of the decay.

Let us calculate now the ratio of the number densities of $X$-particles, to the density of the relativistic species. The latter is taken
as $m_X^3$ at the initial temperature, $T_{in} = m_X$, but in realistic case it can be higher by a factor of few. E.g. the number density
of photons is $0.24 T^3$ and the same is true for other relativistic species: leptons, quarks, gluons, and even for the electroweak bosons, so our assumption leads to some overestimate of $X$ density.
The precise value depends upon the concrete model.    

%\section{Freezing number density}\label{numberdensity}
After freezing, the number density of $X$-particles remains constant in the comoving volume, i.e.:
\be
n_X = n_{Xfr} \left(\frac{ a_{fr}}{a}\right)^3 = n_{Xfr} \left(\frac{t_{fr} }{t}\right)^2 = n_{Xfr} \left(\frac{x_{fr}}{x}\right)^8 , 
\label{nx-of-T}
\ee
where $n_{Xfr}$ is frozen number density of $X$-particles, 
$ a_{fr}$ is the value of the cosmological scale factor at the moment of freezing, and we used the expansion law $a \sim t^{2/3}$ and the relation
between time and temperature $t T^4 =const$.

The energy density of the relativistic particles drops in the course of expansion  from the moment of X-freezing as:
\be
\rho^{rel} = \rho^{rel}_{fr} \left(\frac{t_{fr}}{t}\right) = \rho^{rel}_{fr} \left(\frac{ x_{fr}}{x}\right)^4,
\label{n-rel-of-T}
\ee
where 
\be
\rho^{rel}_{fr}  = \frac{\pi^2 g_* }{30}\,T^4_{fr}
\label{rho-rel-f}
\ee
is the energy density of relativistic matter at the moment of $X$-freezing.

The number density of relativistic particles is related to their energy density according to 
\be
n^{rel} \approx \frac{\rho^{rel}}{3T} = n^{rel}_ {fr} \left(\frac{x_{fr}}{x}\right)^3,
\label{n-rel-of-x}
\ee
where $n^{rel f}  \approx \pi^2 g_* T_{fr}^3/90$.
We neglected here the difference between the effective  number of relativistic species, $g_*$,  in 
the expression for the energy density and the similar coefficient in the expression for the number density. 

Correspondingly:
\be 
\frac{ n_X}{ n^{rel}} = \frac{n_{Xfr} }{n^{rel}_{fr}} \left( \frac{x_{fr}}{x} \right)^5 =
\frac{n_X^{(in)}}{n^{rel}_{ fr}} \left(\frac{x_{fr}}{x}\right)^5   \left(\frac{x_{in}}{x_{fr}}\right)^8 \frac{x_{fr}^5}{Q_f}.
%\frac{\pi^3 (\ln(2Q))^5}{3 \alpha^2 \beta x^5} \,\frac{m_X^3}{m_R m_f^2}
\label{nX-nrel}
\ee
Substituting $Q_f$ from eq.~(\ref{Q-f}) we find:
\be
\frac{ n_X}{ n^{rel}} = \frac{90 \pi}{\alpha^2 \beta_{ann}} \frac{m_X^3}{m_R m_f^2} \left( \frac{x_{fr}}{x} \right)^5 .
\label{nX-nrel-2}
\ee

This ratio would evolve in this way as a function of $x$ till the complete decay of the scalaron at $T=T_{hf}$ (\ref{T-hf}). Using \eqref{x-h-f} and (\ref{x-f-f})  we ultimately find:
\be 
\left(\frac{ n_{X}}{ n^{rel}}\right)_h = \frac{ 90 \pi(\ln(Q_f))^5}{ \alpha^2 \beta_{ann} x^5_{hf}} \,\frac{m_X^3}{m_R m_f^2} =
\frac{90 \pi (\ln(Q_f))^5}{\alpha^2 \beta_{ann} (192\pi^3 g_*)^{5/4}} \, \left(\frac{m_R}{m_{Pl}}\right)^{5/2}
\frac{m_f^3}{m_R m_X^2}.
\label{nX-nrel}
\ee
Later on at  GR stage this ratio does not change much, decreasing only due to the heating of the plasma by the massive particle annihilation.

 As is discussed above, after Eq.~(\ref{f-fin-f}), there could be an additional production of $X$-particles by the decays of
heavier fermions $f$. However, the contribution of such decays into the ratio   $ n_{X}/ n^{rel} $ (\ref{nX-nrel-2})
at the freezing temperature $T_{fr}$ is $T_{fr}/m_R$ and decreases with dropping temperature. The freezing temperature is determined by 
Eq.~\eqref{x-f-f}.
Hence the extra contribution to the number density of X-particles 
can be safely neglected, if 
\be
\frac{90 \pi^2}{\alpha^2 \beta_{ann}} \left(\frac{m_X}{m_f}\right)^2 \ln Q > 1.
\label{X-prod}
\ee
In fact the effect of the additional $X$-production is somewhat weaker due to $f \bar f$-annihilation, which eliminates creation
of $X$-particles. An account of several types of bosonic and fermionic superpartners does not change our estimates if they are
heavier than $X$ roughly by factor ten.

The contemporary energy density of $X$-particle can be approximately estimated as 
\be
\rho_X = m_X n_\gamma \left(\frac{ n_X}{ n_{rel}}\right)_h =  7 \cdot 10^{-9} \frac{m_f^3}{m_X m_R}\, {\rm cm^{-3}},
\label{rho-X}
\ee
where $n_\gamma \approx 412$/cm$^3$ and we take $\alpha = 0.01$, $\beta_{ann} = 10$, $g_*=100$, $m_f = 10^5$ GeV, and $m_X = 10^4$~GeV. 
For the chosen values of the parameters  $ Q_f \approx 1.7 \cdot 10^4$, and $\ln Q_f \approx 10$.
%In fact $g_f < g_*$, so our result is somewhat overestimated.\\

This energy density should be close to the energy density of the cosmological dark matter, $\rho_{DM}~\approx~1$ keV/cm$^3$.
It can be easily achieved with $m_X \sim 10^6$ GeV and $m_f \sim 10^7$ GeV:
\be
\rho_ X = 0.23\,  \left(\frac{m_f}{10^7\, {\rm GeV}}\right)^3 \left(\frac{10^6\, {\rm GeV}}{m_X}\right) \, {\rm \frac{keV}{cm^3}}.
\label{rho-fin}
\ee

\section{Anomalous decay into gauge bosons} \label{s-gauge-anomaly}

Up to now we omitted possible contribution  to the scalaron decay from the the conformal anomaly. As it was 
found in Refs.~\cite{Dolgov:1980kp,Dolgov:1981nw,Dolgov:1993vg}, the coupling of the massless gauge bosons to gravity is determined by the anomaly in the
trace of energy-momentum tensor of the gauge fields which can be presented as:
\be
T_\mu^\mu = \frac{\beta^2 \alpha}{8\pi}\, G_{\mu\nu}  G^{\mu\nu},
\label{trace-anom}
\ee
where $\alpha$ is the fine structure constant, $\beta$ is the first coefficient in perturbative expansion of the beta-function,
and $G^{\mu\nu}$ is the gauge field strength.
Evidently this coupling leads to the decay of the curvature $R(t)$ into gauge bosons.

The decay width of the particle production by curvature due to conformal anomaly, as calculated in
Ref.~\cite{Gorbunov:2012ns}, is
\be
\Gamma_{anom} = \frac{\beta \alpha^2 N}{ 96 \pi^2} \,\frac{m_R^3}{m_{Pl}^2}  . 
\label{Gamma-anom}
\ee
It is to be compared with the decay width into minimally coupled massless scalars:
\be
\Gamma_s = \frac{m^3_R}{48  m_{Pl}^2}.
\label{Gamma-S}
\ee

The former is suppressed by the factor $(\beta \alpha^2 N)/(2 \pi^2)$ but still is much  larger than the decay width into 
fermions \eqref{Gamma-f} with the assumed mass $m_f \sim 10^5 $ GeV. Nevertheless the suppression is
significant and it may allow existence of  very heavy SUSY partners to be dark matter particles. However, a
verification of this hypothesis demands significant calculations and this is a subject of different study.
So we take another route of escape in $N=4$  supersymmetry for which beta-function vanishes and the conformal anomaly is absent, 
see e.g. review \cite{Sohnius:1985qm}, Sec.~13.2, and references therein.

However, $N=4$ super Yang-Mills theories are believed to be unrealistic because they do not allow to introduce chiral 
fermions, even if the symmetry is broken spontaneously. Though spontaneous symmetry breaking is considered  to be the most
appealing way to deal with the theories with broken symmetries, it is not obligatory and the symmetry can be broken 
explicitly. It is possible to break the symmetry "by hand" introducing different masses to particles in the same multiplet.
This would allow to construct a phenomenologically acceptable model. Since the symmetry is broken by mass, the 
theory would remain renormalizable. At higher energies, much larger than the particle masses, it would behave as  $N=4$
super Yang-Mills theory and at this energy scale the trace anomaly would vanish.  

Apart from that there are phenomenologically acceptable $N=1$ and $N=2$ supersymmetric theories which possess the so called 
conformal window, i.e. in this theories with a certain set of the  multiplets  trace anomaly vanishes. For a review and the list of
references see~\cite{Chaichian:2000wr}. 

The conformal coupling of scalar fields to gravity postulated above indeed  breaks supersymmetry but supersymmetry is broken
anyhow and this kind of breaking does not lead to revival of the conformal anomaly.

Gravitational corrections to the trace anomaly leads to appearance of the well known contribution proportional to the 
square of the curvature tensor. This contribution does not lead to production of gauge bosons. Higher loop gravitation
corrections, even if result in gauge boson production, are strongly suppressed.

We thank M. Shifman and A. Vainshtein for the discussion of the presented above features of supersymmetric theories.

\section{Conclusion}

We see that due to the continuous and relatively slow matter production by the oscillating scalaron field, $R(t)$, the temperature of the ordinary
matter in the universe drops much slower than that in the usual FLRW-cosmology. Correspondingly the canonical relation between 
temperature of matter and the cosmological time becomes $T^4 t =  C_{R^2}$, while usually we have $T^2 t = C_{GR} $. Here 
$C_{GR}$ is a universal constant, proportional to the Planck mass, while in $R^2$-cosmology $C_{R^2}$ depends on the model
and may be strongly different for the scalaron decay into bosons or fermions.

In the case of the decays into scalars LSP can play the role of dark matter particle if it is very heavy, noticeably heavier than
the scalaron. In the case of the scalaron decays into fermions or conformally coupled scalars the mass of LSP may be at the level
of $10^3$ TeV, so the situation looks more
promising  from the point of view of  accessibility of LSP in direct experiment.  Still today they are far away from the energy of the
existing accelerators. The search for such dark  matter particles in low background experiments looks presently more feasible.
If they are discovered, it would be an interesting confirmation of $R^2$ inflationary model.\\[2mm]
{\bf Acknowledgements.}\\
The work  of EA and AD was supported by the RSF Grant 19-42-02004. We thank the referee who indicated a problem with the anomalous production of gauge bosons.
\\[3mm]

\end{document}